\documentclass[apj]{emulateapj}
\journalinfo{\textsc{The Astrophysical Journal}}
\slugcomment{Received 2017 April 10; Accepted 2017 September 27}
\usepackage{url}

\usepackage{natbib}
\usepackage{booktabs}
\usepackage{epstopdf}
\usepackage{graphicx}

\usepackage{amsmath}

\shorttitle{Prominence eruption by helical kink instability}
\shortauthors{Vemareddy et al}
\begin{document}
\title{Prominence eruption initiated by helical kink-instability of an embedded flux rope}
\author{P.~Vemareddy$^1$, N.~Gopalswamy$^2$ and B.~Ravindra$^1$}
\affil{$^1$Indian Institute of Astrophysics, Koramangala, Bengaluru-560 034, India}
\affil{$^2$Goddard Space Flight Center, Greenbelt, USA}
\email{vemareddy@iiap.res.in}

\begin{abstract}
We study the triggering mechanism of a limb-prominence eruption and the associated coronal mass ejection near AR 12342 using SDO and LASCO/SOHO observations. The prominence is seen with an embedded flux thread (FT) at one end and bifurcates from the middle to a different footpoint location. The morphological evolution of the FT is similar to an unstable flux rope (FR), which we regard as prominence embedded FR. The FR twist exceeds the critical value. In addition, the morphology of the prominence plasma in 304\AA~images marks the helical nature of the magnetic skeleton with a total of 2.96 turns along arc length. The potential field extrapolation model indicates that the critical height of the background magnetic field gradient falls within the inner corona (105Mm) consistent with the extent of coronal plasma loops. These results suggest that the helical kink instability in the embedded FR caused the slow rise of the prominence to a height of the torus instability domain. Moreover, the differential emission measure analysis unveils heating of the prominence plasma to coronal temperatures during eruption, suggesting a reconnection-related heating underneath the upward rising embedded FR. The prominence starts with a slow rise motion of 10km/s, followed by fast and slow acceleration phases having an average acceleration of $28.9m/s^2$, $2.4m/s^2$ in C2, C3 field of view respectively. As predicted by previous numerical simulations, the observed synchronous kinematic profiles of the CME leading edge and the core supports the involved FR instability in the prominence initiation. 
\end{abstract}

\section{Introduction}
\label{Intro}
Solar prominences are the most common coronal features, with plasma embedded in the magnetic environment lying along and above magnetic inversion lines \citep{hanssen1974, hanssen1998}. They contain cool and dense plasma indicating thermally and pressure isolated from the surrounding corona. They appear at the limb as bright features when observed in optical and EUV lines, and in microwaves; they appear dark, referred to as filaments, on the disk. Of the many scientific aspects of prominences,  explaining their loss of equilibrium to erupt as part of coronal mass ejections (CMEs) is one of the major aspects relevant to space-weather \citep{VialEngvold2015}. 
 
The first prominence model, proposed by \citet{kippenhahn1957}, assumes a magnetic configuration where the gravity force of the prominence is balanced by the Lorentz force.  This basic model has now evolved into sheared arcade and flux rope \citep{antiochos1998, antiochos1999, pneuman1983, rust1996,amari2003a} models. In both models, the prominence material rests in the dipped regions of magnetic field lines \citep{mackay2010}. Accordingly, breakout \citep{antiochos1999} and tether-cutting  \citep{moore2001,moore2006} reconnection mechanisms in the sheared arcade scenario, helical kink and torus-instabilities \citep{forbes1991, priest2002, torok2005, kliem2006,zhangj2012} in the flux rope scenario have been proposed and employed to explain the underlying physical mechanism in most of the observed prominence/filament eruptions. While reconnection plays a fundamental role in the earlier models, ideal MHD instability leads to the onset of eruption in the later models.  

High quality space based multi-wavelength observations from the Transition Region and Coronal Explorer (TRACE; \citealt{handy1999} ) and Atmospheric Imaging Assembly (AIA; \citealt{lemen2012}) on board the \textit{Solar Dynamics Observatory} (SDO; \citealt{pesnell2012}) have helped testing the proposed models with the observations. Tether-cutting reconnection has been found to play a triggering role in several observational studies \citep[e.g.,][]{liuchang2007,yurchyshyn2006a, liur2010, vemareddy2012a}. A reconnection in the overarching loop system, which is identified commonly as a brightening, sets in the run-away tether-cutting reconnection below the sheared arcade, and subsequent eruption of core field as CME. The CME initiation by helical kink-instability of flux rope is evidenced in the recent observational studies (e.g., \citealt{williamsdr2005, alexanderd2006, srivastavaak2010, pankaj2012}). Signatures of kinking-writhing of flux rope like structure are substantiated by the observed morphology in different wavelengths \citep{green2007, gilbert2007}. In fact, local distribution of photospheric magnetic twist supports the origin of the underlying twist \citep{vemareddy2014b, vemareddy2016b}. The role of torus instability becomes significant after the flux rope reached a certain height, from where steeply decreasing magnetic field gradient in the ambient corona drives the eruption \citep{torok2004}. Recent studies found a characteristic curve of horizontal field gradients with height for eruptive flares as an indication of the ability of the flux rope to experience torus instability \citep{chengx2011}.

Most observational studies claim direct evidence of ideal kink-instability with the flux system underlying the prominence body (e.g., \citealt{rust1996,  srivastavaak2010, pankaj2012}).  Sometimes  prominences appear as kinked illuminating in hot EUV channels. These are interpreted as twisted flux ropes, which are uncommon \citep{lites2005, zhangj2012,chengxin2014}.  Large-scale prominence structures, filled with cool dense plasma, had been noticed  with long-time stability, even over months \citep{mackay2010}. Such cases require a different mechanism for stability loss and the subsequent eruption. Therefore, the global structure and equilibrium of the prominence environment is an open question.

Recent observational studies find that unstable part of the filament belongs to a part of the magnetic structure that the filament is hosted \citep{liur2012,vemareddy2014b}. Reconciling the different structures observed in filaments and prominences on the disk versus limb is crucial for understanding the triggering mechanisms of eruptions. With this background, we studied an erupting prominence that occurred on May 9, 2015, showing evidence for the prominence-associated flux rope undergoing ideal kink instability and causing the large-scale CME eruption.  Such cases are uncommon as the instability part is generally the prominence itself and identifying it in on-disk observations is crucial in identifying the trigger mechanism. For completeness, we also studied the prominence thermal properties and the kinematics of the associated CME. A brief description of the observational data is given in Section~\ref{obs}, the analysis results are presented in Section~\ref{res} and Summary and discussion are given in Section~\ref{disc}. 
\begin{figure*}[!htp]
\centering
\includegraphics[width=.99\textwidth,clip=]{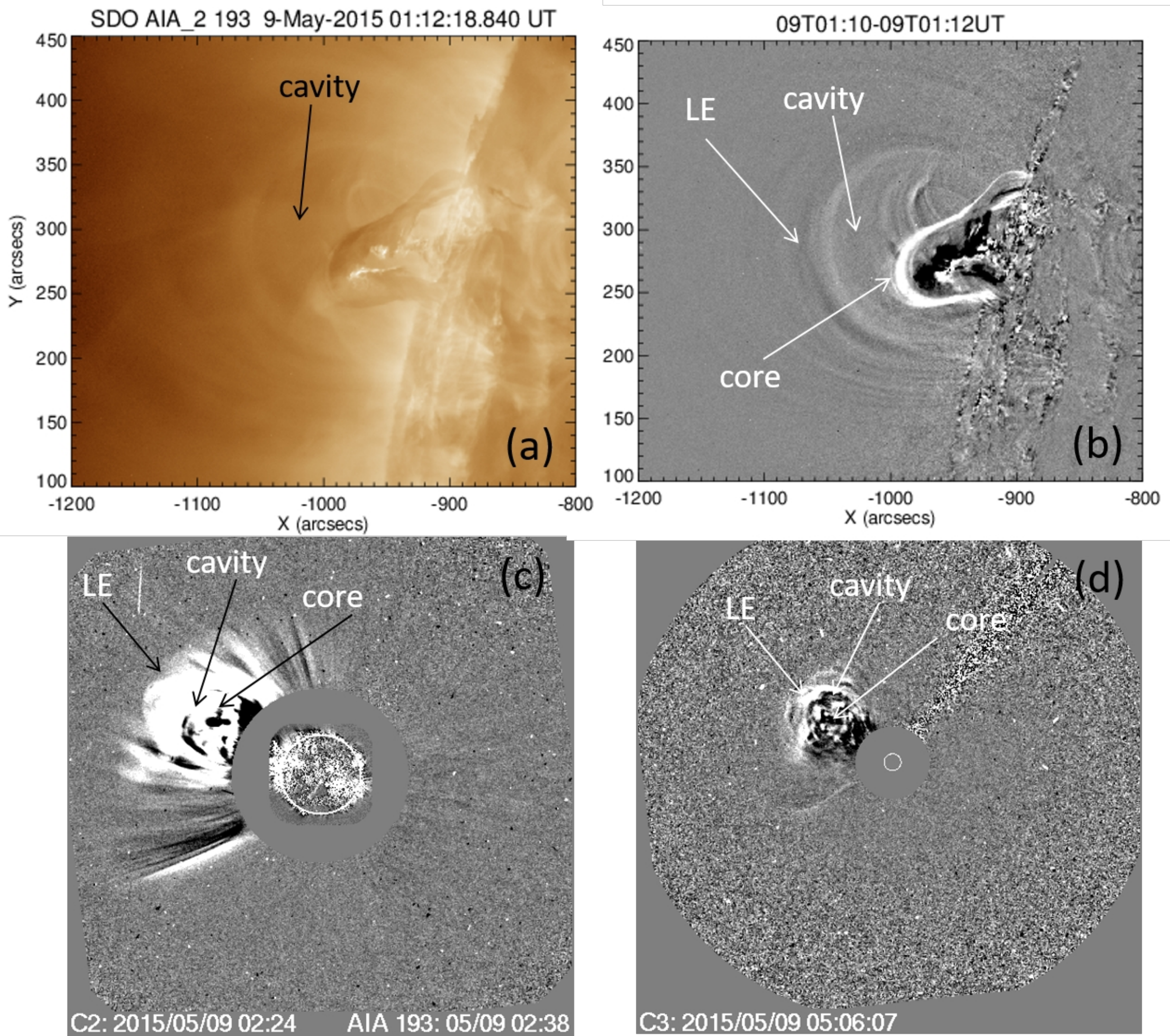}
\caption{Observations of prominence eruption to CME. a) AIA 193 snapshot of prominence being erupting, b) difference image in 193\AA~showing clear three-part structure of core, cavity and leading edge, c) and d) erupted prominence being observed in LASCO C2 and C3 as white-light CME. }
\label{Fig1}
\end{figure*}

\section{Observations}
\label{obs}
We use observations from the SDO and the Large Angle and Spectrometric Coronagraph (LASCO, \citealt{brueckner1995})  on board the Solar and Heliospheric Observatory (SoHO). The dynamics of the prominence, and post eruption loops upto 1.2$R_\odot$ are captured in the multi-thermal EUV images of the AIA at a cadence of 12s. The magnetic roots of the coronal plasma structures are studied using photospheric line-of-sight magnetograms from the Helioseismic and Magnetic Imager (HMI; \citealt{schou2012}). HMI obtains full disk line-of-sight magnetograms at a cadence of 45s and vector magnetic fields at a cadence of 135s. The CME evolution in the extended outer corona is imaged by white-light coronagraphs LASCO/C2 (1.5-6$R_\odot$) and LASCO/C3 (3.5-32$R_\odot$). In addition, we obtained GONG H$\alpha$ images in H {\sc i} 6563\AA~wavelength at 1$^{\prime\prime}$ per pixel resolution.

\begin{figure*}[!ht]
\centering
\includegraphics[width=.99\textwidth,clip=]{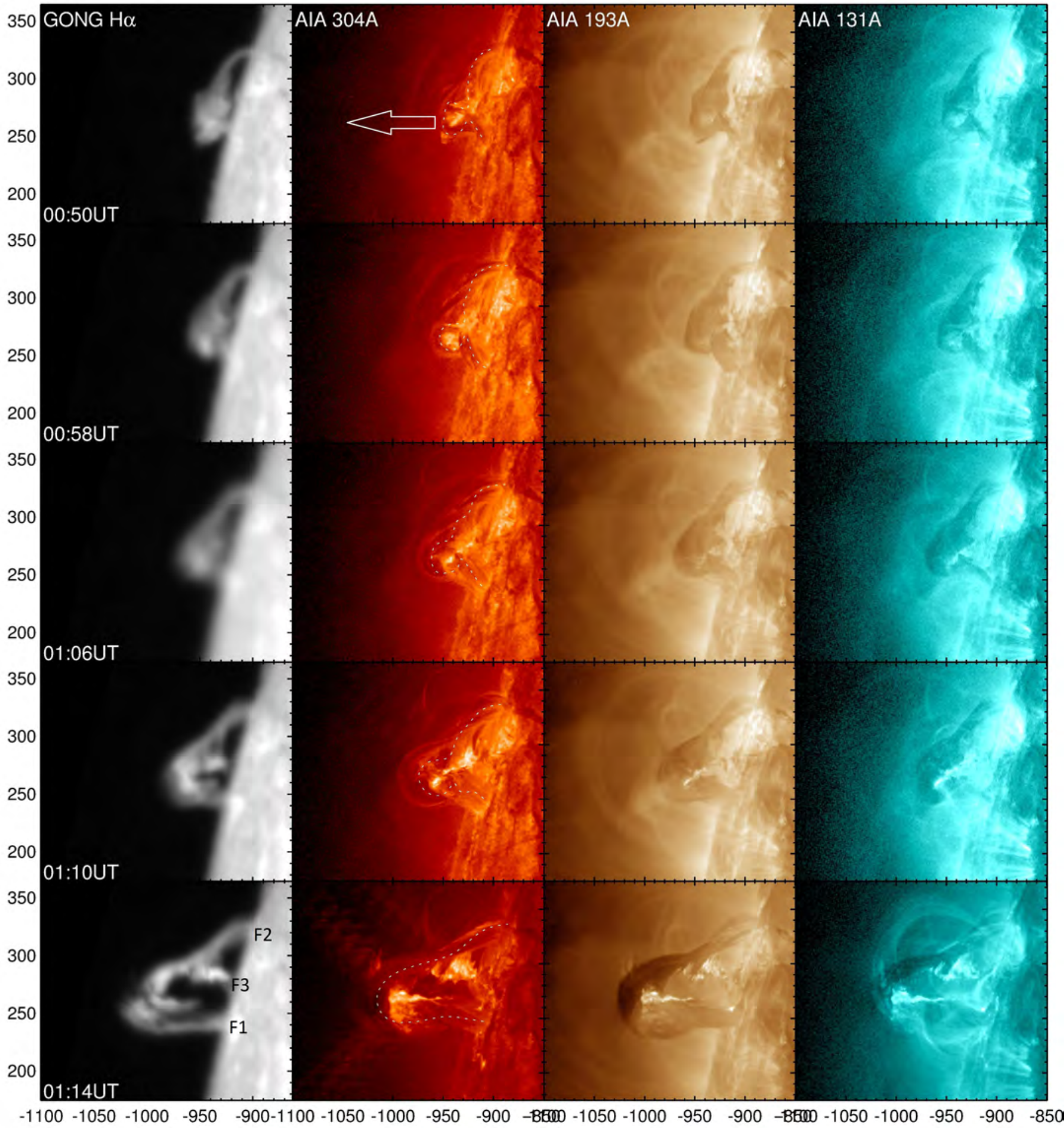}
\caption{Initiation of prominence eruption on May 9, 2015. First Column: Observations from GONG H$\alpha$  Second, third, fourth columns: observations in AIA 304, 193, 131\AA~pass bands. The chromospheric connections of prominence legs (F1, F2, F3) are clearly distinguishable in H$\alpha$ 01:14UT panel. In addition to core prominence, the AIA 193, 131\AA~images also delineate the entrapping closed field environment. All panels are in heliographic coordinate system in arc second units. }
\label{Fig2}
\end{figure*}

The prominence erupted on May 9, 2015, at 01:00UT from NOAA active region (AR) 12342 located near the east limb (N18E53). Figure~\ref{Fig1} presents the erupting prominence in the AIA 193\AA~pass band and the white-light LASCO observations of the CME. Time difference maps in AIA 193\AA~show the extended over-arching loops and the surrounding less bright cavity region together enveloping the core prominence oriented approximately in north-south direction. When the prominence reached the field-of-view of the LASCO/C2, it was found to be in the interior part of the CME preserving the three part structure \citep{hundhausen1987}  up to a distance of approximately $15R_\odot$. Its further propagation in LASCO/C3 FOV follows merged and diffused structure with dominant lateral expansion than radial motion.   

\section{Analysis and Results}
\label{res}
\subsection{Morphology}

In Figure~\ref{Fig2}, we plotted the summary of a multitude of observations showing the morphological evolution of the prominence during the onset of the eruption (00:00--01:14UT on May 9). H$\alpha$ observations show the prominence and the roots of its legs in the lower chromosphere. The barb section, as a third leg, bifurcates from the apex to a different footpoint location (F3) in the lower chromosphere, and it is visible while the prominence rises. Movies reveal more details on the supporting magnetic structure in the prominence system. There are essentially two branches of flux threads intermingled as a single structure toward the southern footpoint F1. While rising, the flux threads from the leg of F3 appear to have a different branch than the main prominence. Recent studies using AIA data interpret two different flux systems stacked over each other where the lower one, in the form of sigmoid, sets to erupt \citep{vemareddy2012a, liur2012,vemareddy2014b} which is in question in different cases. The May 9 event is an example to such cases having two branches of flux threads that are connected to each other at one end. The visibility of any flux system depends on the thermal conditions of the embedded plasma against the disk.  

The AIA 304\AA~images, which give plasma information from the chromosphere and transition region (He~{\sc ii} $\sim10^4$ K), show the prominence and its surroundings filled with plasma. In both H$\alpha$ and 304\AA, the prominence contains bright emitting plasma. In hotter channels of 193, 211, 131\AA, the prominence plasma is opaque (optically thick) and the coronal closed field environment overly the prominence. From simple scaling measurement, we infer that the entrapping field extends upto 110Mm above the photosphere. Since the peak temperature sensitivity is $>10^6$K in 131\AA, the prominence material appears dim compared to the ambient plasma loops.  Unlike the H$\alpha$, EUV observations provide more details on the curvature of the prominence. Its lower section, before the onset, is curved and dipped down possibly due to the helical nature of the magnetic field (dotted curve). Such structures appear to have S or inverse-S shape when seen in projection onto the solar surface. From the observed curvature, we infer an S-shaped prominence. 
\begin{figure*}[!ht]
	\centering
	\includegraphics[width=.99\textwidth,clip=]{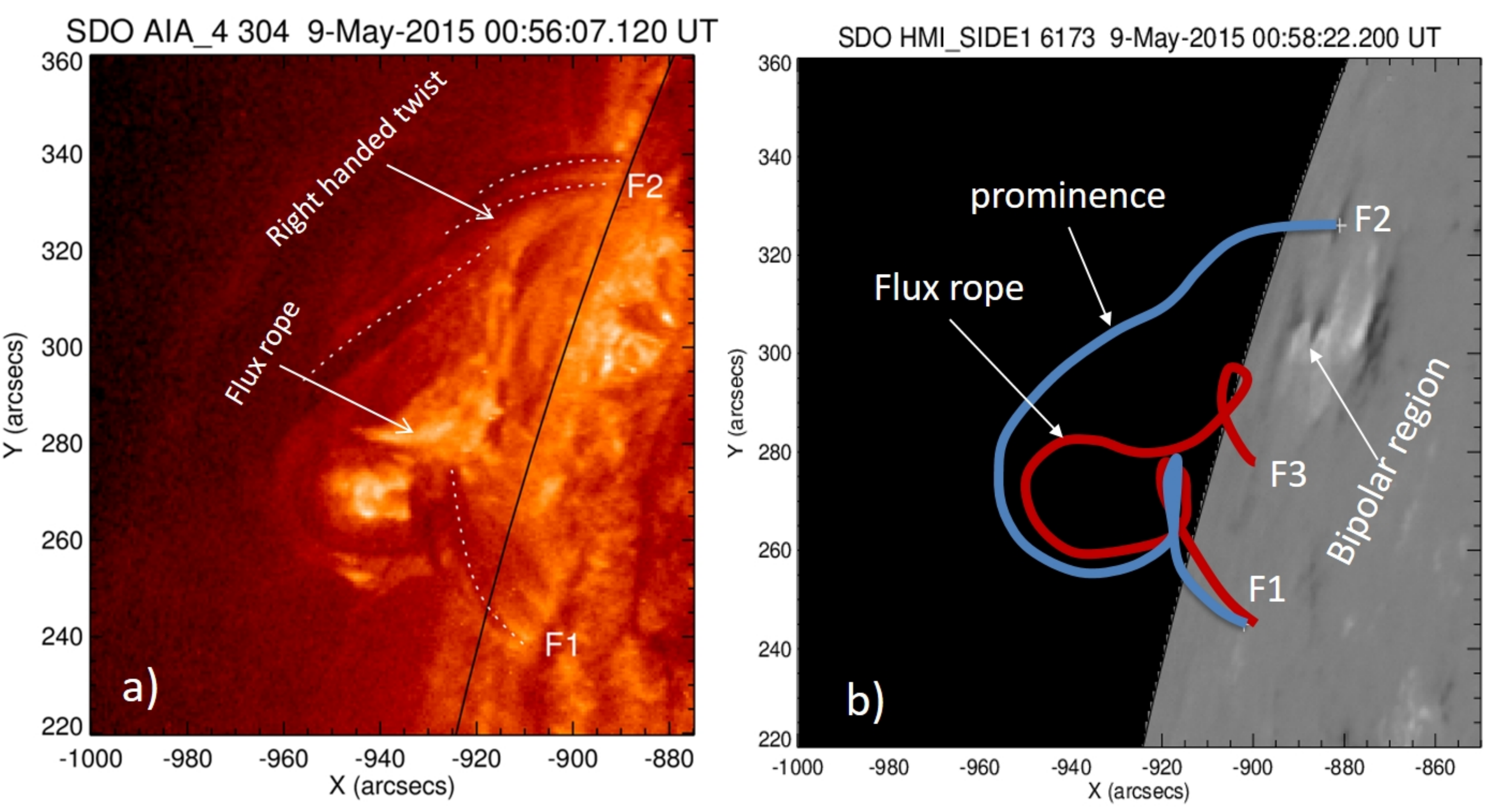}
	\caption{ a) Evidence for right handed twist in the supported magnetic field of prominence. Dotted curves trace dark lanes between twisted magnetic flux threads, providing helicalness of magnetic field. Also, as seen from F1, the entire magnetic structure winded in right-hand, a signature of magnetic writhe, b) Trace of both prominence and the embedded MFR are plotted on LOS magnetogram. Footpoints F1, F2, F3 are rooted in north, south, south polarity, respectively. }
	\label{Fig3_1}
\end{figure*}
Flare ribbons are formed in the chromosphere due to the coronal reconnection under the rising prominence, and indeed have S-shape ribbon morphology as seen in UV (see \texttt{AIA\_1600\_211.mp4}) and H$\alpha$ observations. In tandem, post eruption arcade (PEA) is seen in EUV images straddling the polarity inversion line (PIL, see Figure~\ref{Fig9}). However, the PEA is too weak to be recognized in disk-integrated GOES X-ray flux. As the AR magnetic flux associated with footpoints is seen to be diffusing in time, it is likely that the magnetic structure supporting the prominence is formed by slow magnetic reconnection  \citep{pneuman1983, vanballegooijen1989, amari2003a}) over days before this eruption.

\subsection{Helical kink nature}
The morphology of the prominence in 304\AA~marks the helical nature of the magnetic field lines and their handedness. When the magnetic threads are twisted in a bundle, the gap between the threads appears as a dark lane around the bundle compared to the thread in 304\AA emission. The trace of either dark or bright feature essentially provides the twisted nature of magnetic field in the prominence as can be clearly seen in 304\AA~images (dotted curves). Being clockwise skewed, the field lines (threads) are twisted in right hand direction with positive helicity (Figure~\ref{Fig3_1})a. Note that field line helicalness is present well before the eruption but not as a consequence of the eruption. Therefore, the system of prominence and the enveloping stabilizing field supports the flux rope models (e.g., \citealt{forbes1991, priest2002,roussev2003, torok2005,kliem2006,zhangj2012}). Further, as seen from footpoint F1, the entire prominence structure is wound in clockwise direction, which is a signature of magnetic writhe arising from excess twist. 

After a careful examination of 304\AA~movie, we sketch the trace of the prominence and the embedded FR in Figure~\ref{Fig3_1}b for a possible magnetic configuration. The background image HMI LOS magnetogram shows the magnetic connections of the prominence structure. Due to projection, the flux distribution appears very weak with a field strength less than 400G. The prominence appears to be located in the periphery of the AR 12342 containing major negative polarity sunspot.  Overall, the bipolar distribution shows PIL along the projected length of the prominence. The prominence footpoints F1 and F2 correspond to positive and negative polarity respectively. As discussed earlier, the FR is intermigled with prominence structure towards F1 and deviates from apex towards negative polarity region F3. Being inclined to the surface, the geometrical structure of FR is ambiguous and indicates a low lying structure following curved PIL below it. 

\begin{figure*}[!ht]
	\includegraphics[width=.9\textwidth,clip=]{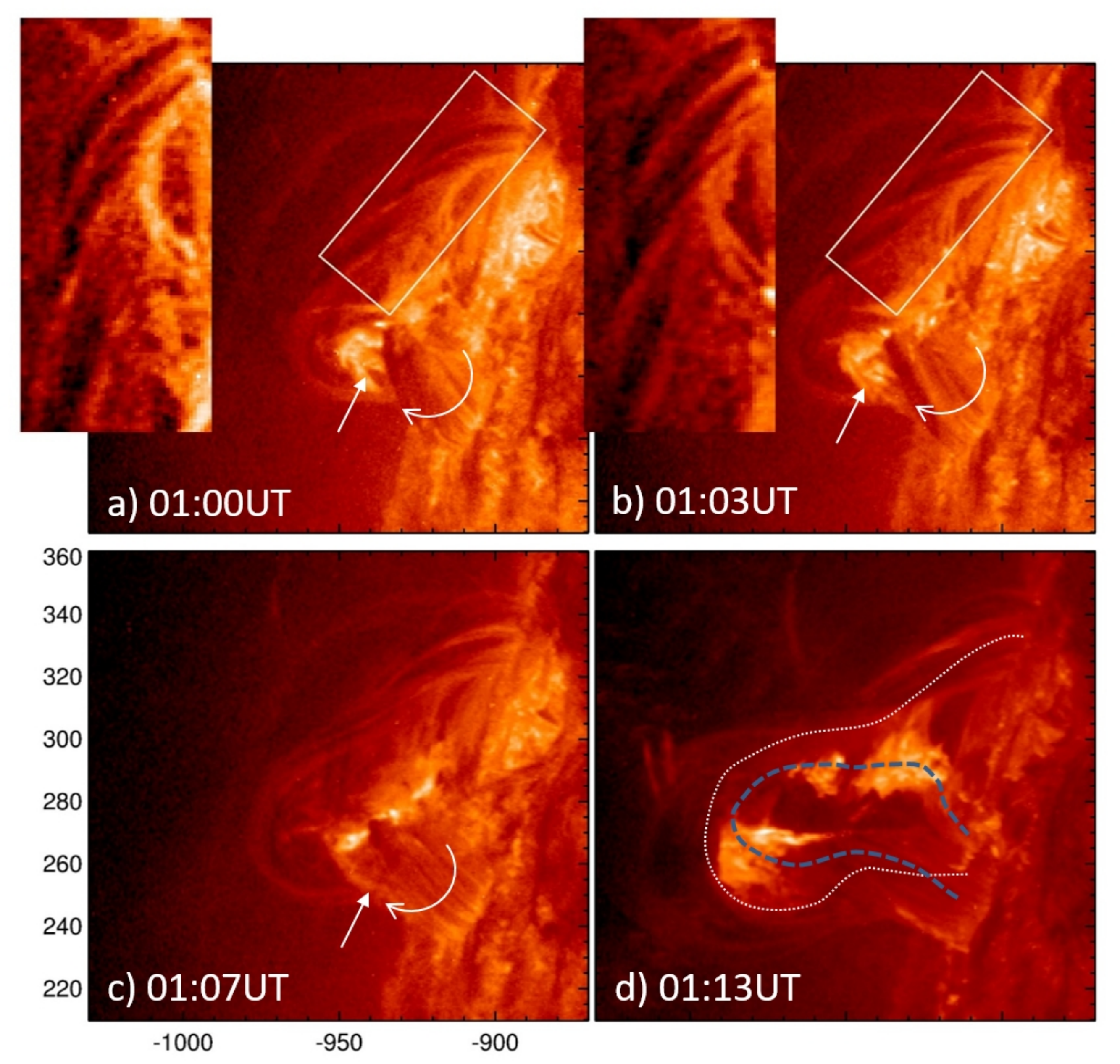}
	\caption{Signatures of magnetic twist evolution during slow rise motion. In a) and b) rectangular region is zoomed in the inset to show the right-helical threads in the prominence. Arrows point to writhed section of the prominence including MFR and curved arrows indicate rolling motion (clockwise from above) of fieldlines resulting from conversion of writhe to twist during slow upward rise motion. In panel (d), traces of prominence and MFR are indicated with dotted, dashed curves.  (Also see movie \texttt{AIA\_304.mp4})}
	\label{Fig3_2}
\end{figure*}

As depicted in Figure~\ref{Fig3_2}, the FR exhibits a dynamical evolution associated with the brightenings, pushing the prominence upward and subsequently becomes an isolated structure from a different footpoint location between F1 and F2. While rising, both the prominence and the FR stretch in length and the prominence exhibits a rolling motion which is associated with the clockwise rotation motion of the magnetic structure towards F1. A section of the prominence is zoomed in the inset to clearly indicate the presence of the right-helical threads. Since the footpoints of the prominence structure remain attached to the surface (line-tied), the observed rotation of magnetic structure is purely a consequence of transformation from (positive) writhe to (positive) twist of the fieldlines. Other possibilities, as discussed in \citet{suyingna2013}, related to twist increase by reconnection in the shared arcade, and/or untwisting of magnetic field in the rising flux rope during expansion and relaxation can be ruled out because the observed conditions are not supportive. Alternatively, we point that the rotation of magnetic structure is also reported to be driven by vortex motion of footpoints, which is clearly not the case here, efficiently injecting twist into the coronal magnetic structure  \citep{bonet2010,wedemeyerbohm2012,vemareddy2012b}. 

 \begin{figure*}
	  	\centering
	  	\includegraphics[width=.89\textwidth,clip=]{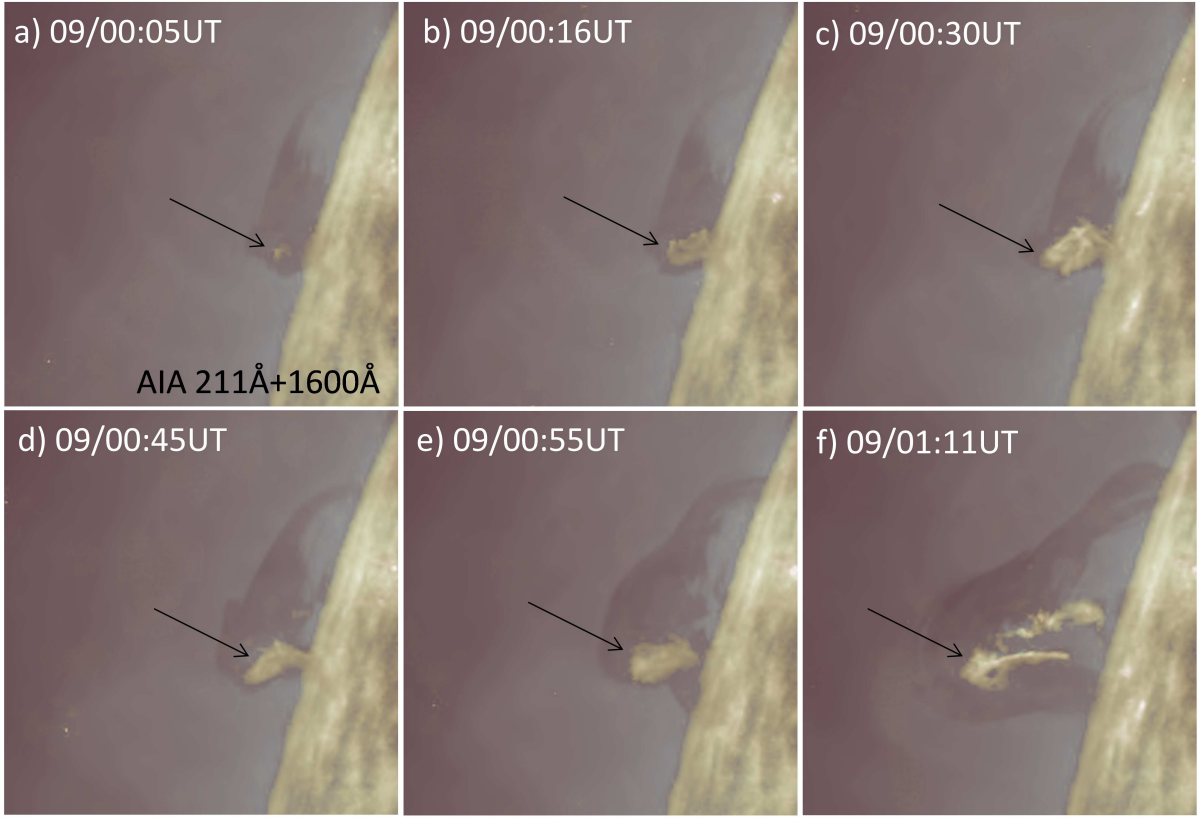}
	  	\caption{Blended images of prominence in AIA 211 and 1600\AA~wavelength bands. Arrows point to the flux thread (1600\AA) embedding the cool prominence (211\AA) at one end and bifurcates from middle to a different footpint location. Corresponding to the dynamic activity of this flux thread in time, similar to a kink unstable flux rope, the entire prominence structure sets to slow upward motion. (also see movie \texttt{AIA\_1600\_211.mp4})}
	  	\label{Fig4}
 \end{figure*}
	  
From these 304\AA~images, we infer the half pitch length ($\pi p$) of most observed helical tracers as 27Mm. And the arc length (L) of the prominence at the time of its fast rise (09/01:00UT, before expansion) is measured as 160Mm, which is a typical value found in statistical studies \citep{wangyuming2010}. With these observables, we deduce the number of turns ($N=\frac{L}{2\pi p}$, see \citealt{srivastavaak2010}) that the field lines in the prominence body have is 2.96 ($\sim6\pi$). Because of this preexisted high twist number, the entire prominence body is kinked (deformed), but still remains in stable equilibrium presumably by the downward gravity force of the entrapped mass and a strong strapping effect of the over-arching loops (Figure~\ref{Fig1}b). 


\subsection{Initiation mechanism}
Initiation of the prominence rise commences from 09/00:00UT. The key information about the onset mechanism comes from AIA 1600\AA~(C IV) images, which capture plasma emission from the upper photosphere and transition region peaking at $LogT[K]=5.0$ \citep{lemen2012}. In Figure~\ref{Fig4}, we plotted the blended (different proportions of transparencies) AIA 1600 and 211\AA~images at different epochs of the evolution. They clearly show different thermal conditions of the plasma across the prominence body. Compact bright flux threads belonging to FR are illuminated in AIA 1600\AA~images whereas the surrounding twisted flux threads, belonging to the main prominence (see also Figure~\ref{Fig3_1}), are in a cool plasma environment captured in 211\AA~ images. Similar to the recent notion that more twisted flux threads are illuminated in hot passbands \citep{zhangj2012}, it implies that the flux threads from F3 are more compactly twisted than those from F2. We regard this compact flux thread (arrows in Figure~\ref{Fig4}) as a distinct flux rope, embedded in the large scale prominence.        

The compact flux thread undergoes helical deformation during the early activation period 00:00UT-00:40UT as revealed by AIA 1600\AA~ images. The movie gives an impression of right hand rotation of the apex, although the handedness of field line twist is not clear. Note that the 1600\AA emission originates from around the central axis of the prominence due to projection. During this period, the enveloping prominence rises slowly with a registered outlier distorted shape in EUV observations.  We interpret this dynamical evolution of the embedded flux thread (or FR) as due to the helical kink instability, since the magnetic twist exceeds the critical value \citep{torok2004}, and hence triggers the onset of prominence slow rise.  Importantly, the helical deformation related to the kink instability in the FR needs to uplift the overlying prominence, in addition to the entire closed field environment. Accounting for the smaller bulkiness of the FR compared to the prominence, the FR instability leads only to a slow upward lifting of the prominence over an hour. Note that the field lines enveloping the flux thread (belongs to main prominence) are also helical with right handed (positive) twist, which are already at the theoretical critical value of the twist. As a result, the entire prominence body is kinked (helically deformed) appearing as S-shape morphology in projection. It is worth note that the exact amount of twist required for a kink instability depends on several factors, including the loop geometry and neighboring/overlying fields, but is generally agreed to be at least one full wind \citep{hood1979,baty2001,torok2004}. 

In many well-observed cases, flux systems with twists exceeding the critical value undergo kinking and writhing (see also the movie), with right (left) handed twist resulting in clockwise (anti-clockwise) apex rotation \citep{gilbert2007,alexanderd2006}. It is important to point that the helical instability arises in FR with indications of apex rotation that is not very clear in observations. With the observed dynamical activity of the FR, the prominence rise is initiated from a preexisting marginal stability state (also see Discussion section). With the onset of the rise motion, the prominence exhibits rolling motion from FP2 to FP1. As a consequence, the writhe (helical deformation of the axis) converts to fieldline twist (helicalness of the threads) as per conservation of helicity. This writhe to twist transformation is clearly indicated by the clockwise rotation motion of the magnetic structure in the prominence leg at F1 (Figure~\ref{Fig3_2}). To preserve the sign of the net helicity content, the clockwise (positive) writhe gives right-hand twist, which we see as the clockwise motion of the magnetic structure. We note the reconnection-related heating prevails mostly with the FR, which also supports our presumption of two separate branches of flux systems. 
	  \begin{figure*}
	\centering
	\includegraphics[width=.99\textwidth,clip=]{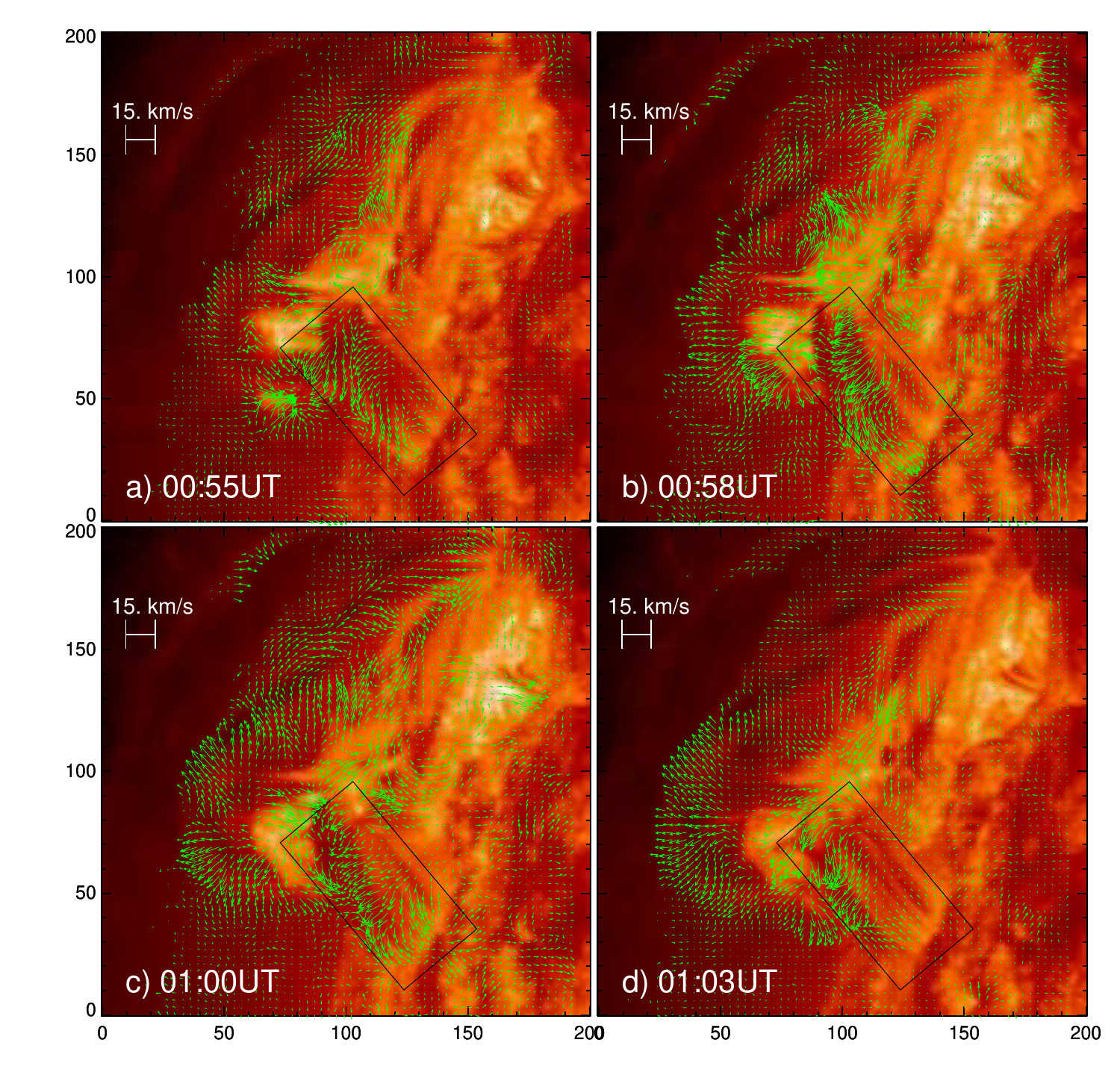}
	\caption{Plasma flow velocity (arrows) in the prominence. Organized flow pattern in the south leg (rectangular box ) corresponds to rotating motion, whereas those in the apex part corresponds to a net upward motion. Maximum length of arrows scales to 15km/s. Axis units are in pixels of 0.6 arcsec size.}
	\label{Fig5}
\end{figure*}

\begin{figure*}
	\centering
	\includegraphics[width=.99\textwidth,clip=]{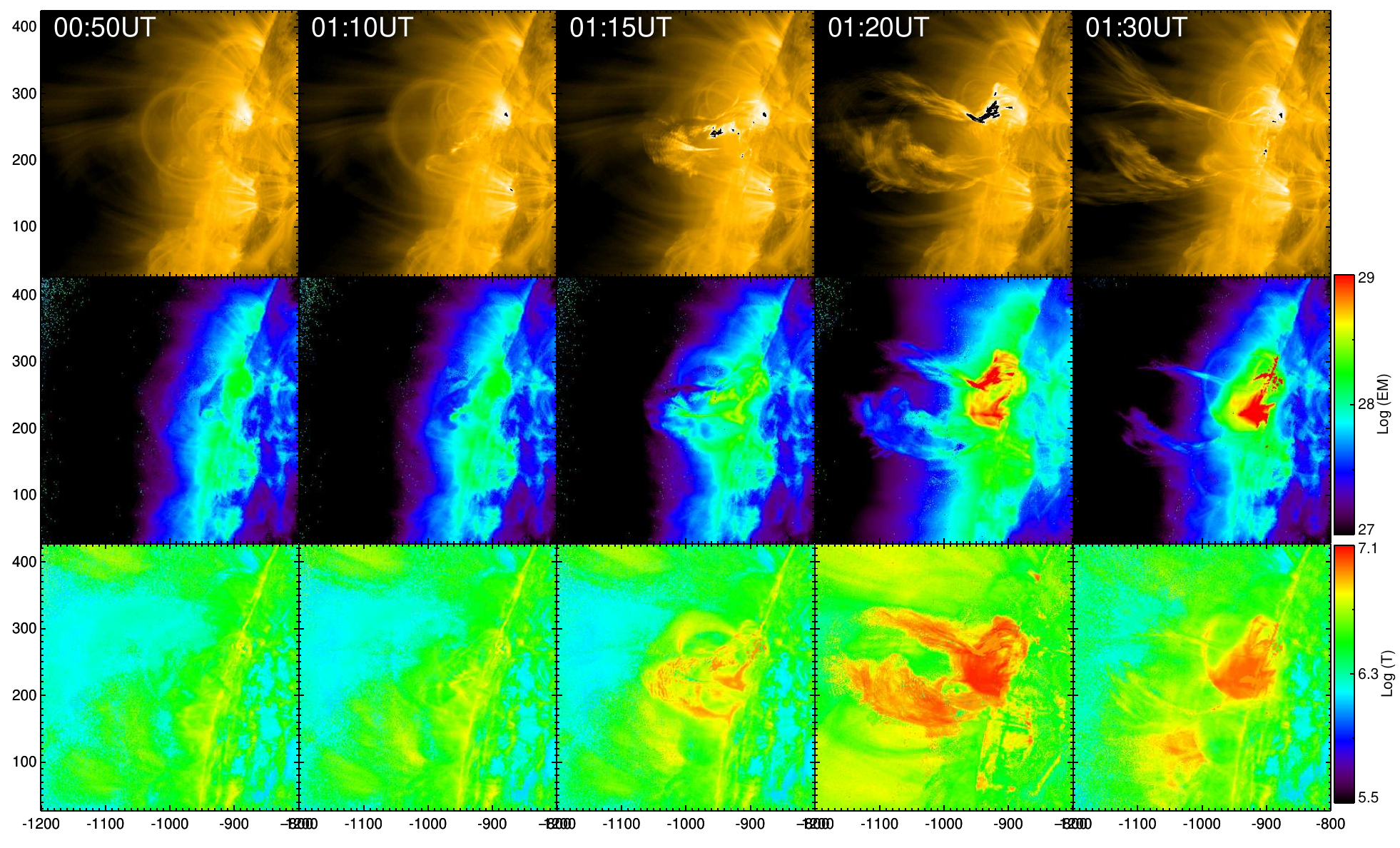}
	\caption{ DEM analysis during the prominence eruption top row: AIA 171\AA~ observations showing the erupting prominence, middle row: The corresponding EM maps, bottom row: average temperature ($\bar{T}$) maps in Logarithmic scale. Note that the heated prominence body (upto 10MK) visible in better contrast from the background in temperature maps. Panels at 1:20 \& 1:30UT also show the PEA under the rising prominence. Axes units are in arc seconds in heliographic coordinate system.}
	\label{Fig6}
\end{figure*}

\subsection{Plasma flow in the prominence}
To derive plasma flow in the prominence structure, we used the Differential Affine Velocity Estimator (DAVE; \citealt{schuck2005}) method on AIA 304\AA~ images of one minute cadence. The method works on optical flow principle to detect transverse flow velocity of features accounting for contraction, dilation, and/or rotation. The procedure involves induction equation expansion satisfying both convection and advection models for a magnetofluid. In our case, a window of 15 pixels with advection is used to derive the flow velocity of the plasma. 

In Figure~\ref{Fig5}, we  show the plasma flow velocity on AIA 304\AA~images during the prominence slow rise. In different parts of the prominence, the flow velocity exceeds 15km/s  with different flow patterns, which is of the order of slow rise velocity. The flow vectors in the prominence apex obviously indicate a net upward motion of the prominence. Organized flow vectors in the south leg, indicated by a rectangular box, corresponds to clockwise rotating motion as discussed earlier. The rotation motion is a process redistributing the twist of the field lines along the kinked prominence body. The derived plasma flow vectors captured this process remarkably during the rolling and slow rise motions. The plasma drains down along the threads in south leg as vector direction points towards surface.  In order to capture further detailed plasma flow along the helical threads, high resolution images are required, which are averaged in the tracking window of AIA images. 

\subsection{Thermal properties}
The thermal structure of the prominence is studied with a differential emission measure (DEM) analysis using images in the six EUV channels of AIA. We used \texttt{xrt\_dem\_iterative2.pro}, available in Solar Soft Ware (SSW; \citealp{freeland1998}) package, with modifications \citep{golub2004, weber2004}. The code implements a forward fitting procedure to construct DEM at each pixel given its flux and temperature response function in each passband. Similar to earlier studies \citep{chengx2012, vemareddy2014b}, we constructed DEM maps of the prominence structure and evaluated DEM-weighted average temperature ($\bar{T}$) and thermal emission measure (EM) defined as

\begin{equation}
\bar{T}=\tfrac{\int{DEM(T)TdT}}{\int{DEM(T)\,dT}};\,\,\,\,EM=\int{DEM(T)\,dT}
\end{equation}

Integrations are evaluated over the temperature range of $5.6 < log T < 7.3$, {which excludes the typical prominence temperature of $LogT=4$ }. The temperature and EM maps together with AIA 171\AA~observations are shown in Figure~\ref{Fig6}. The images are scaled appropriately for a better contrast from the background emission. Till 09/01:10UT, the EM and $\bar{T}$ remain in orders of 27 and 6 respectively, as soon as activation ensues, both EM and T increase by an order. We can notice the heated prominence body with a better contrast (by an order) from the background.  The diameter of the prominence body before activation is 30Mm. Considering a mean EM  value of $5\times10^{27}cm^{-5}$ along the prominence body, the plasma density ($n=\sqrt(EM/l)$) comes out to be $1.29\times10^9cm^{-3}$. 

\begin{figure*}
	\centering
	\includegraphics[width=.99\textwidth,clip=]{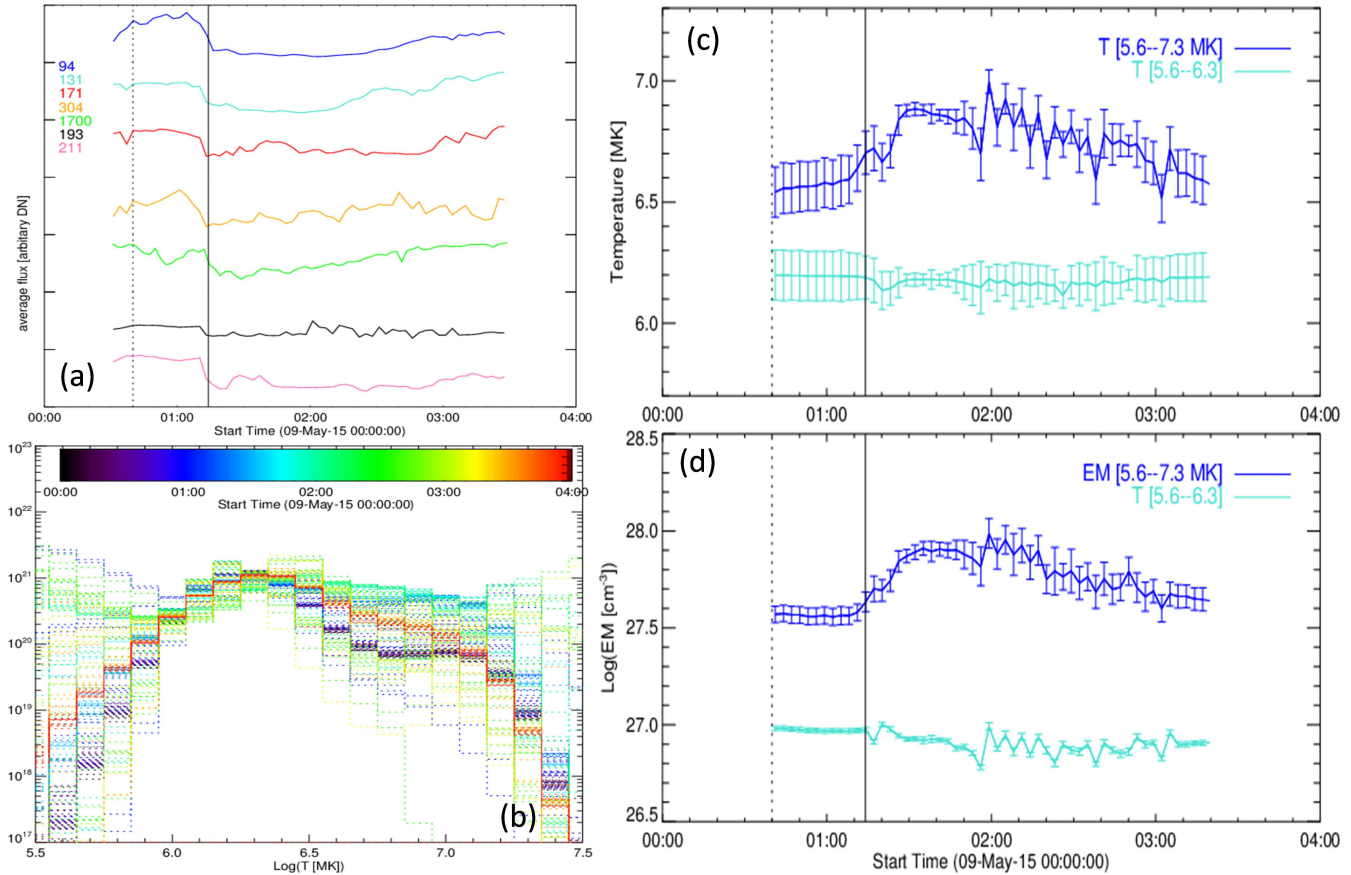}
	\caption{a) Light curves from the field-of-view of the prominence. All AIA channels show a fall in the intensity as a reason of emission from rising cool prominence compared to pre-conditions. b) Variation of DEM in time. DEM curves are color scaled on time range. DEM is well constrained in the temperature range $5.6<Log T (K)<7.3$, c) and d) Temperature and emission measure integrated in the temperature ranges of $5.6<Log T (K)<7.3$ and $5.6<Log T (K)<6.3$, respectively.  Error bar is standard deviation of the resulted parameter after hundred montecarlo estimations.   }
	\label{Fig7}
\end{figure*}

In order to study the time evolution, we further derived DEM for the averaged intensity in 6 wavelength band observations of the prominence at one-minute intervals. To reduce the contribution of the background emission, the chosen area minimally covers the prominence in its entire rise motion. The light curves of this average intensity are plotted in Figure~\ref{Fig7}(a). From these light curves of different wavebands, it is clear that the background emission is dominant in the early rise phase (before 01:15UT) compared to the later phase when the prominence body expands to a maximum extent. Correspondingly, the light curves show fall in their intensity after 1:15UT in most AIA channels. Note that the intensities are normalized to unity in each pass band, and there exists excess emission in hot channels compared to that of cold channels. It is indeed the case after the prominence reached a certain height. In the flux rope scenario, the reconnection in the thinning current sheet underneath rising fluxrope (FR) is the source of heating (see \texttt{AIA\_304.mp4}).  Then the prominence appears in hot channels due to the heated plasma emission and the light curves recover from the dip after the onset of the eruption. 

The temperature dependence of DEM over time is plotted in Figure~\ref{Fig7}(b) (see also \citealt{sunjq2014}). The emission measure over the entire FOV varies in the range $10^{17-22}cm^{-5}$. Its variation is about the mean temperature of 6.3MK, with major emission from low temperatures in the early initiation phase, and dominant emission from high temperatures after the onset. This also confirms that the DEM is well constrained in the chosen temperature range for valid observables (also see \citealt{chengx2011,vemareddy2014b})

In Figure~\ref{Fig7}(c-d), we plot EM and $\bar{T}$ derived by integrating the DEM in two temperature bands. One is full temperature range $5.6<Log T(K)<7.3$ and the other is $5.6<Log T(K)<6.3$ excluding the hotter emission. The later band is chosen to distinguish the prominence from the background. Errors are deduced after 100 monte-carlo iteration runs of fitting by introducing errors to the input observations and are shown by vertical bars. Over the entire temperature range, both $\bar{T}$ and EM show gradual increase(LogT:6.52-6.7, Log(EM):27.6-27.7) since the onset of upward rise (00:40UT), which is followed by a relatively high increase (up to factors of 6.9, 28) after the impulsive rise motion (1:15UT). The fact that we see enhanced EM and $\bar{T}$ while the light curves in each band show sudden dip after 1:15UT is due to a relatively intense flux from hotter bands (193, 94, 131\AA) compared to the low temperature bands. It is worth noting that after 1:20UT, the emission from PEA contributes significantly to $\bar{T}$  and EM. However, in the low temperature range, EM and $\bar{T}$ varies about a mean values, and a dip immediately after onset corresponds to dip in light curves related to the dominant cool emission from the prominence body. Since the sampling bands are hotter than $LogT=5.6$, the emission corresponds to the prominence-corona transition region rather than the interior of the prominence body. In summary, the rising prominence and the FR are heated by progressive reconnection upto 10MK. 

\subsection{Kinematic evolution}
\begin{figure*}
\centering
\includegraphics[width=.96\textwidth,clip=]{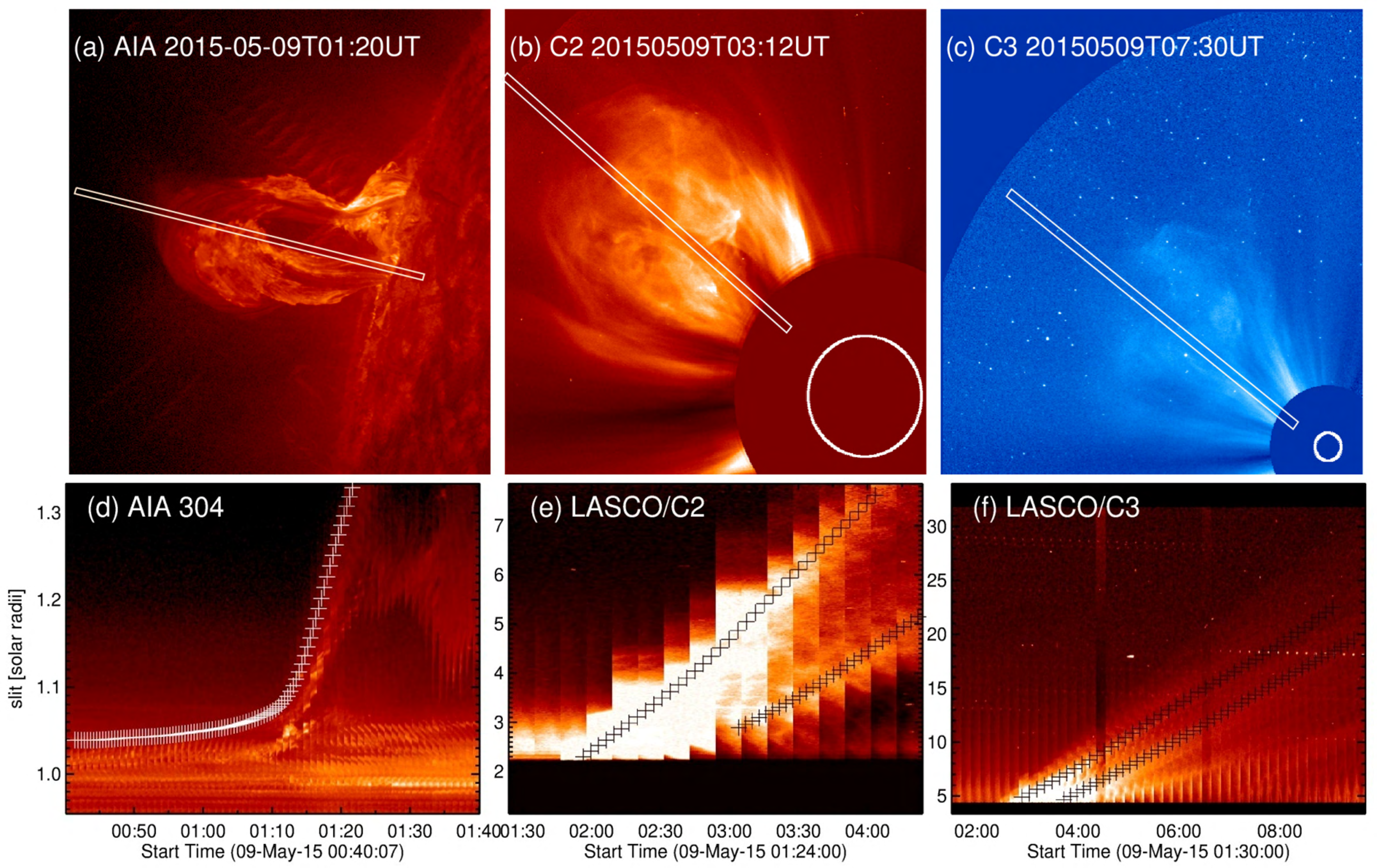}
\includegraphics[width=.72\textwidth,clip=]{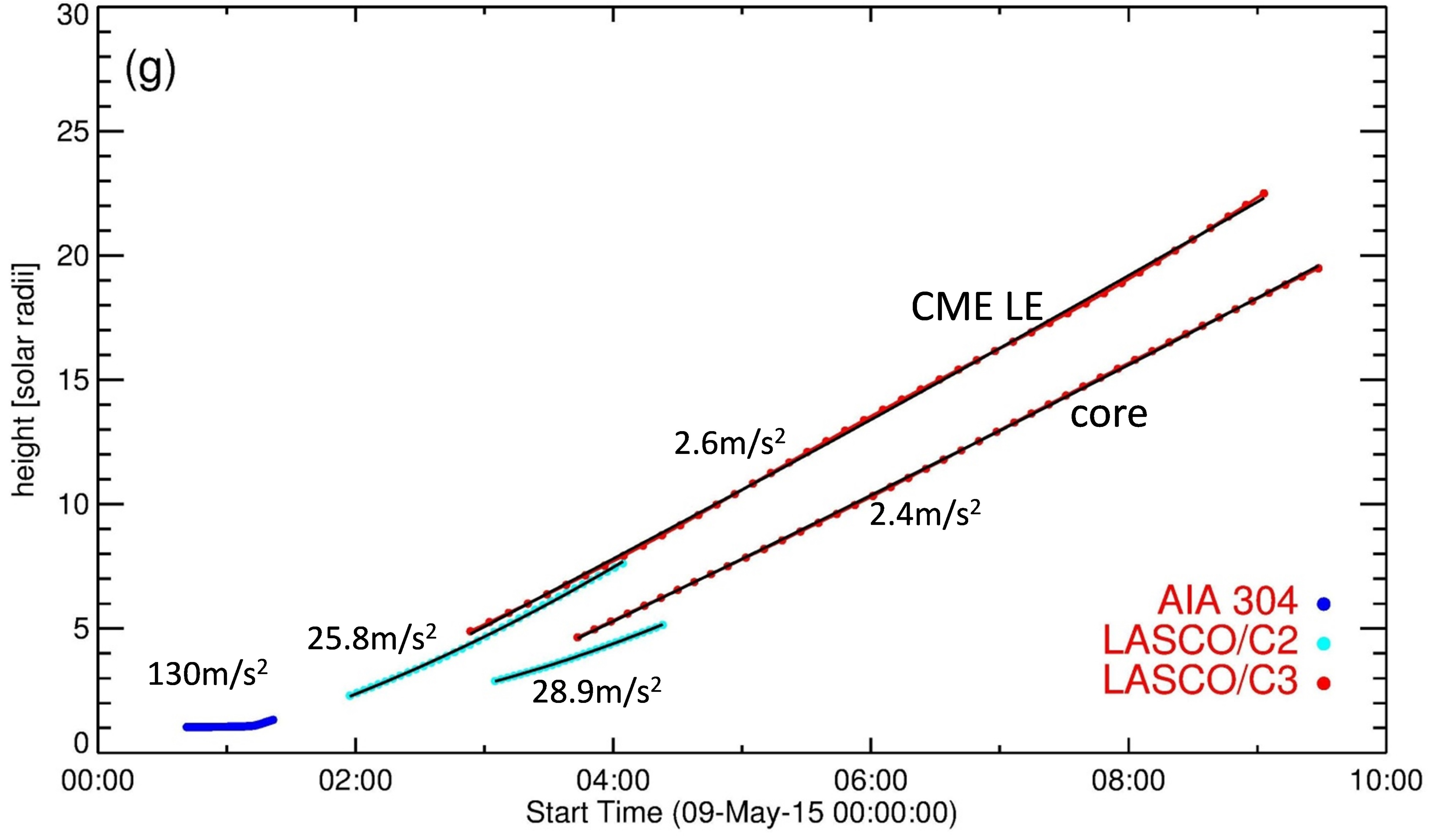}
\caption{ Kinematics of the prominence and its associated CME. a-c) Slit positions across the prominence in AIA, and the CME in LASCO/C2, LASCO/C3 observations, d-f) space-time stack plots of the slit obtained from AIA, C2, C3 observations. Trace of the prominence, the CME leading edge, and core is shown with ``+” in respective panels. g) Height-time measurements of the prominence (blue dots), the CME leading edge and the core. Cyan (red) dots corresponds to height time history in C2 (C3) observation. The data points are fitted with second order polynomial and the derived average acceleration is marked near the curves. }
	\label{Fig8}
\end{figure*}

The erupted prominence appeared in the LASCO/C2 field-of-view at 09/01:36UT and further in LASCO/C3 at 09/02:30UT. Since it is a limb event, the projection effects have less influence in the derived kinematic parameters. We derive the kinematic properties by stacking the rectangular slit images obtained over the CME/prominence evolution. In Figure~\ref{Fig8}(a-c), we show the rectangular slits placed across the prominence/CME apex. Their respective stack images are plotted in Figure~\ref{Fig8}(d-f). We then identified the bright feature corresponding to the leading edge (LE) of the prominence in AIA and the LE of the CME, core of the prominence separately in LASCO observations \citep{gopalswamy2003}. In C2 stack plot, separate traces clearly correspond to the LE and core. However, in the C3 stack plot, both of these traces are visible in the early period upto 04:30UT and become faint gradually. We carefully followed these traces corresponding to the core and LE and measured the position from the sun center for height time history. The height-time history of the CME is also given in the SOHO/LASCO CME catalog \citep{gopalswamy2009}. The LE had an average speed of 661 km/s within the combined C2-C3 FOV\footnote{\url{ https://cdaw.gsfc.nasa.gov/CME\_list/UNIVERSAL/2015\_05/htpng/20150509.013617.p066g.htp.html}}. and an average deceleration, -2.65 $m/s^{2}$.

In Figure~\ref{Fig8}(g),  we show the composite height-time plot of the CME/prominence from AIA to LASCO/C3 fields-of-view. Data points are fitted with spline smoothing procedure to minimize spikes in the measurements. The velocity and acceleration are then derived. From them, we show average acceleration value near the curves in the same panel. During the activation period, the prominence rises very slowly having a speed of 10-150km/s, which corresponds to acceleration upto 0.9$km/s^2$. In this phase, the prominence has an average acceleration of 130m/s$^2$ which essentially includes a slow average acceleration of 18m/s$^2$ upto 01:10UT before the fast rise. 

In the LASCO FOV, both the CME LE and the prominence core show nearly synchronous kinematic profiles, indicating that the different structures in the CME are moving together. They contain fast and slow acceleration stages in C2 and C3 height ranges. The LE attained a peak velocity of 588km/s in C2 FOV, which was slowed slightly to 573km/s in C3 FOV. However, this corresponds to significant difference in average acceleration (25.8m/$s^2$, 2.6m/s$^2$). The CME core has greater peak velocity of 573km/s in C3 than in C2 of 523km/s.  It has slightly higher average acceleration (28.9m/$s^2$) than CME LE. Thus in the outer coronagraph FOV, the CME continued to accelerate as the case of Mar 11, 2012 CME studied in \citet{gopalswamy2015}. Thus the kinematic study shows three stages of acceleration, slow, fast and slow consistent with the general population of prominence associated CMEs \citep{gopalswamy2003}.

The synchronous kinematic profiles of CME LE and core sheds some insights on the role of triggering mechanism on CME kinematics. The models that invoke FR instability found a synchronized motion of CME LE and the prominence \citep{torok2005} compared to those explained by resistive instabilities like breakout or tether-cutting models. For example, in a detailed study of a simulated breakout CME, \citet{karpen2012} found that the prominence-carrying portion of the structure moves at Alfvenic speeds during the classic impulsive phase of the CME/eruptive flare, which then travels more slowly than the CME front. This picture of kinematic properties is not obvious in an observational study of 18 CMEs \citep{maricic2009}. Therefore, the identified synchronous trajectory in our case also supports the FR instability scenario.


\section{Summary and Discussion}
\label{disc}
This work presents a detailed study of a prominence eruption identifying the global magnetic structure and the location of instability that led to the eruption. The analysis shows that the identified instability occurs in a flux thread embedded in the prominence, despite both are joined at one end. The AIA 1600\AA~observations are the key to identify the preeruptive dynamical activity of the flux thread in the prominence channel. The FR is bifurcated at the apex of the prominence to a different footpoint location. Such cases are discovered in recent studies, from the continuous AIA observations, as double decker filaments \citep{vemareddy2012a, liur2012, vemareddy2014b} when observed against the disk. Depending on different evolving conditions, one of the branches will be subjected to instability. In the case presented in \citet{vemareddy2014b}, the lower branch (although difficult to determine in projection against the disk) is inferred to have FR characteristics, and initiated to erupt by kink-instability. In the present case, the onset mechanism is the helical kink-instability that arose in the inner flux system as an embedded FR of the observed large scale prominence. 
\begin{figure}[!ht]
\centering
\includegraphics[width=.49\textwidth,clip=]{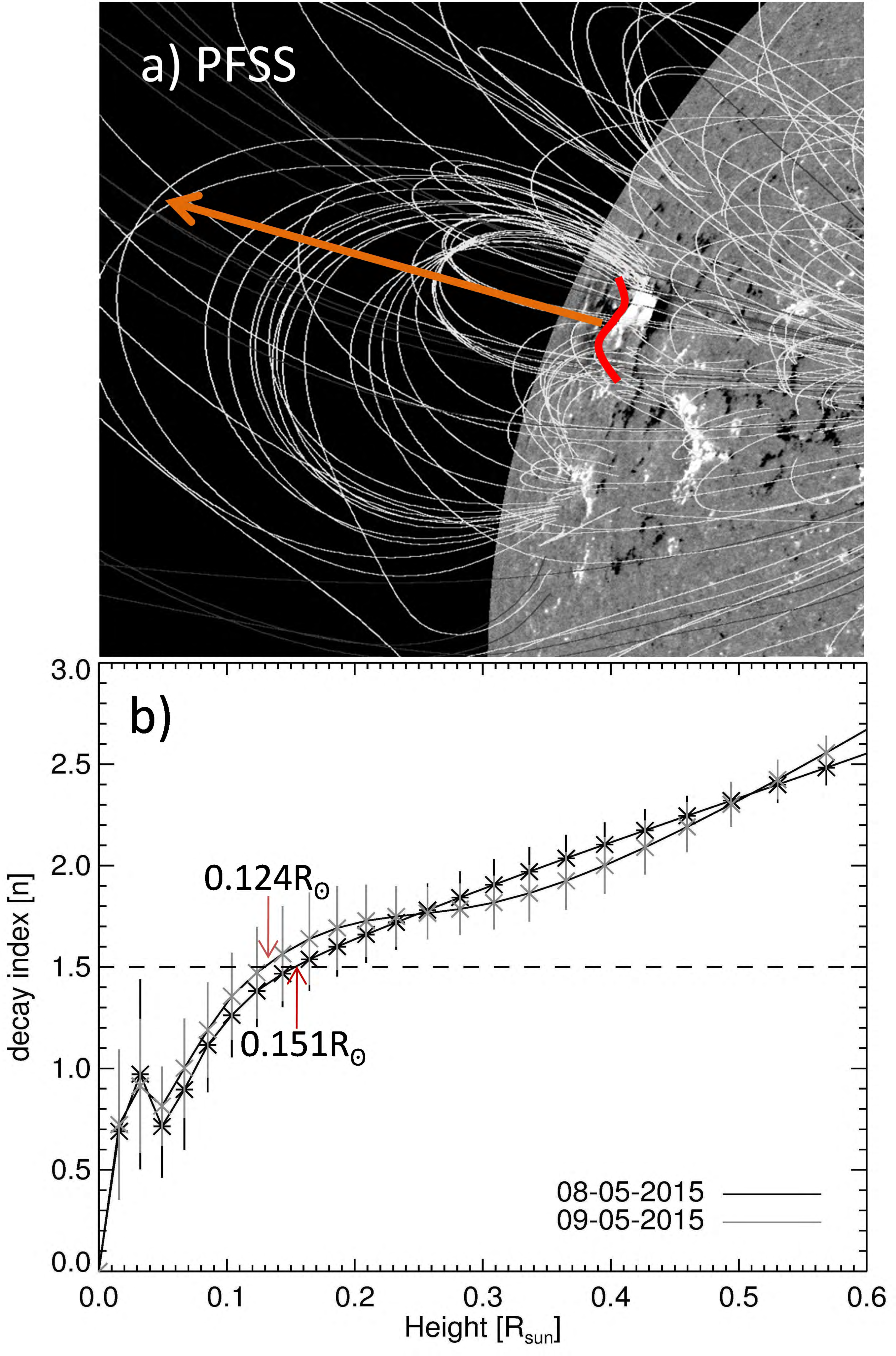}
\caption{Decay index of background coronal magnetic field above the prominence a) Field lines of PFSS extrapolation on LOS magnetogram. Red curve represents the PIL underneath the prominence, arrow points to vertical direction of background field gradients being computed, b) Decay index $n$ as a function of height from the solar surface. The curve approaches $n_{crit}$  at 0.124, 0.151$R_{\odot}$ on May 8, 9 respectively. Error bars are standard deviation of $n$ obtained at 10 points along the PIL. }
	\label{Fig9}
\end{figure}
The DEM analysis characterizes the thermal emission and temperature structure of the prominence-core transition region. EM varies about mean temperature of 6.3MK, with major emission from low temperature in the early initiation phase, and dominant emission from high temperature after the onset, suggesting reconnection related heating in the later phase. 

The associated CME moves at an average velocity of 545km/s in LASCO FOV, which is higher than the average value (475km/s) of the general population of CMEs \citep{gopalswamy2003}. Starting from an initial speed of 10km/s, the CME moves away in slow, fast and slow acceleration phases. The CME LE and core have average accelerations (25.8m/s$^2$, 28.9m/s$^2$) in C2 FOV. As a typical behaviour of most CMEs associated with prominences, the CME continues to accelerate in the outer coronagraphic FOV.  


In order to reveal the role of the background magnetic field in driving the eruption, we calculate the decay index $n=-\frac{d\log ({{B}_{h}})}{d\log h}$ \citep{kliem2006} of the horizontal field using potential field source surface (PFSS) code available in the SSW package. The code takes into account the evolving field on the full sphere by assimilating magnetograms into a flux-dispersal model \citep{schrijver2003}, and yields the coronal field in spherical coordinate system. After converting the field to Cartesian coordinates, we calculate n at ten points along the PIL below the prominence and plotted in Figure~\ref{Fig9} as a function of height from the solar surface. The error bars are standard deviation of n at those 10 points. The curve shows a bump at around 0.03$R_\odot$ (20Mm), indicating that the transverse field in the low corona decreases rapidly enough to allow an eruption to occur, which is consistent with eruptive flares for which the bump appears at 10Mm in instantaneous local magnetograms \citep{chengx2011,vemareddy2014b}. 

For the magnetograms of May 8 and 9, the theoretical threshold of the torus instability, $n_{crit}= 1.5$ corresponds to a height of 0.124, $0.151R_\odot$ respectively ($sim$105Mm). Recall that the observed height of the extended closed field coronal environment is 110Mm (Figure~\ref{Fig1}b). The dip in the curve (20-40Mm) can also be understood in terms of the coronal field configuration and its strength in constraining the prominence eruption. Recent studies point to the significant role of low-altitude field strength, and asymmetric field configuration in the confinement of prominence/filaments \citep{liuy2009}. A study by \citet{liuy2008} reports that the low-altitude field strength for failed eruptions is three times stronger than that for full eruption cases. In the present case, the prominence is lying horizontal to the surface and experiences strong confinement by the low-lying loops along the PIL, causing very slow rise motion of the prominence in the early phase despite being with critically twisted threads. This is how the prominence achieves near stability during 00:40UT-01:10UT even after introducing the instability with a large twist and is likely the regime of failed eruptions of kink-unstable filaments as indicated in \citet{liuy2008}. After crossing this elevated point, the prominence stretches vertically with material draining down, which accommodates the growth of the instability in the field of  symmetric extended coronal loops up to 105Mm.  Altogether, the confinement of the background field is in agreement with the height time (Figure~\ref{Fig8}) plot showing the slow upward motion of the prominence up to 105Mm (1:15UT) in AIA field of view. This is the height range in which the prominence rises under the influence of the kink-instability against the background field.  

As soon as the prominence crosses this height, the self force drives further eruption because the rapidly decaying closed field can no more suppress the rise motion. Recent observations indicate that the critical height for torus stability is as high as 236Mm \citep{wangw2017}, that in order to suppress the further eruption triggered by kink instability. For a confined eruption studied by \citet{guoy2010}, the $n$ lies always below 101Mm before reaching the $n_{crit}$, suggesting a stronger restraining field over the flux rope. While there is a dividing line of $n$ for failed and successful eruptions, the height of $n_{crit}$ depends on background field strength which is different for different cases. Also depending on the magnitude of erupting core field related to hoop/self force, confined/ejective eruption is realised. For example, a kink-unstable filament ascent is terminated within a projected height of 80Mm to a failed eruption following M2.2 flare \citep{haishengji2003}, however a similar kink-unstable filament from AR 10696 rises to about 70Mm which subsequently manifests as a fast CME \citep{williamsdr2005}. 

\acknowledgements SDO is a mission of NASA's Living With a Star Program. SOHO is a project of international cooperation between ESA and NASA. This work utilizes data obtained by the Global Oscillation Network Group (GONG) Program, managed by the National Solar Observatory, which is operated by AURA, Inc., under a cooperative agreement with the National Science Foundation. We thank the referee for insightful comments and suggestions. P.V. is supported by an INSPIRE grant of AORC scheme under the Department of Science and Technology.  The work of NG was supported by NASA's Heliophysics Guest Investigator program.

\bibliographystyle{apj}

\begin{thebibliography}{66}
	\expandafter\ifx\csname natexlab\endcsname\relax\def\natexlab#1{#1}\fi
	
	\bibitem[{{Alexander} {et~al.}(2006){Alexander}, {Liu}, \&
		{Gilbert}}]{alexanderd2006}
	{Alexander}, D., {Liu}, R., \& {Gilbert}, H.~R. 2006, \apj, 653, 719
	
	\bibitem[{{Amari} {et~al.}(2003){Amari}, {Luciani}, {Aly}, {Mikic}, \&
		{Linker}}]{amari2003a}
	{Amari}, T., {Luciani}, J.~F., {Aly}, J.~J., {Mikic}, Z., \& {Linker}, J. 2003,
	\apj, 585, 1073
	
	\bibitem[{{Antiochos}(1998)}]{antiochos1998}
	{Antiochos}, S.~K. 1998, \apjl, 502, L181
	
	\bibitem[{{Antiochos} {et~al.}(1999){Antiochos}, {DeVore}, \&
		{Klimchuk}}]{antiochos1999}
	{Antiochos}, S.~K., {DeVore}, C.~R., \& {Klimchuk}, J.~A. 1999, \apj, 510, 485
	
	\bibitem[{{Baty}(2001)}]{baty2001}
	{Baty}, H. 2001, \aap, 367, 321
	
	\bibitem[{{Bonet} {et~al.}(2010){Bonet}, {M{\'a}rquez}, {S{\'a}nchez Almeida},
		\& {et al}}]{bonet2010}
	{Bonet}, J.~A., {M{\'a}rquez}, I., {S{\'a}nchez Almeida}, J., \& {et al}. 2010,
	\apjl, 723, L139
	
	\bibitem[{{Brueckner} {et~al.}(1995){Brueckner}, {Howard}, {Koomen}, \& {et
			al}}]{brueckner1995}
	{Brueckner}, G.~E., {Howard}, R.~A., {Koomen}, M.~J., \& {et al}. 1995,
	\solphys, 162, 357
	
	\bibitem[{{Cheng} {et~al.}(2014){Cheng}, {Ding}, {Zhang}, {Srivastava}, {Guo},
		{Chen}, \& {Sun}}]{chengxin2014}
	{Cheng}, X., {Ding}, M.~D., {Zhang}, J., {Srivastava}, A.~K., {Guo}, Y.,
	{Chen}, P.~F., \& {Sun}, J.~Q. 2014, \apjl, 789, L35
	
	\bibitem[{{Cheng} {et~al.}(2011){Cheng}, {Zhang}, {Ding}, {Guo}, \&
		{Su}}]{chengx2011}
	{Cheng}, X., {Zhang}, J., {Ding}, M.~D., {Guo}, Y., \& {Su}, J.~T. 2011, \apj,
	732, 87
	
	\bibitem[{{Cheng} {et~al.}(2012){Cheng}, {Zhang}, {Saar}, \&
		{Ding}}]{chengx2012}
	{Cheng}, X., {Zhang}, J., {Saar}, S.~H., \& {Ding}, M.~D. 2012, \apj, 761, 62
	
	\bibitem[{{Forbes} \& {Isenberg}(1991)}]{forbes1991}
	{Forbes}, T.~G., \& {Isenberg}, P.~A. 1991, \apj, 373, 294
	
	\bibitem[{{Freeland} \& {Handy}(1998)}]{freeland1998}
	{Freeland}, S.~L., \& {Handy}, B.~N. 1998, \solphys, 182, 497
	
	\bibitem[{{Gilbert} {et~al.}(2007){Gilbert}, {Alexander}, \&
		{Liu}}]{gilbert2007}
	{Gilbert}, H.~R., {Alexander}, D., \& {Liu}, R. 2007, \solphys, 245, 287
	
	\bibitem[{{Golub} {et~al.}(2004){Golub}, {Deluca}, {Sette}, \&
		{Weber}}]{golub2004}
	{Golub}, L., {Deluca}, E.~E., {Sette}, A., \& {Weber}, M. 2004, in , 217
	
	\bibitem[{{Gopalswamy} {et~al.}(2003){Gopalswamy}, {Shimojo}, {Lu}, {Yashiro},
		{Shibasaki}, \& {Howard}}]{gopalswamy2003}
	{Gopalswamy}, N., {Shimojo}, M., {Lu}, W., {Yashiro}, S., {Shibasaki}, K., \&
	{Howard}, R.~A. 2003, \apj, 586, 562
	
	\bibitem[{{Gopalswamy} {et~al.}(2015){Gopalswamy}, {Yashiro}, \&
		{Akiyama}}]{gopalswamy2015}
	{Gopalswamy}, N., {Yashiro}, S., \& {Akiyama}, S. 2015, \apj, 809, 106
	
	\bibitem[{{Gopalswamy} {et~al.}(2009){Gopalswamy}, {Yashiro}, {Michalek},
		{Stenborg}, {Vourlidas}, {Freeland}, \& {Howard}}]{gopalswamy2009}
	{Gopalswamy}, N., {Yashiro}, S., {Michalek}, G., {Stenborg}, G., {Vourlidas},
	A., {Freeland}, S., \& {Howard}, R. 2009, Earth Moon and Planets, 104, 295
	
	\bibitem[{{Green} {et~al.}(2007){Green}, {Kliem}, {T{\"o}r{\"o}k}, {van
			Driel-Gesztelyi}, \& {Attrill}}]{green2007}
	{Green}, L.~M., {Kliem}, B., {T{\"o}r{\"o}k}, T., {van Driel-Gesztelyi}, L., \&
	{Attrill}, G.~D.~R. 2007, \solphys, 246, 365
	
	\bibitem[{{Guo} {et~al.}(2010){Guo}, {Ding}, {Schmieder}, {Li},
		{T{\"o}r{\"o}k}, \& {Wiegelmann}}]{guoy2010}
	{Guo}, Y., {Ding}, M.~D., {Schmieder}, B., {Li}, H., {T{\"o}r{\"o}k}, T., \&
	{Wiegelmann}, T. 2010, \apjl, 725, L38
	
	\bibitem[{{Handy} {et~al.}(1999){Handy}, {Acton}, {Kankelborg}, {Wolfson}, \&
		{et al}}]{handy1999}
	{Handy}, B.~N., {Acton}, L.~W., {Kankelborg}, C.~C., {Wolfson}, C.~J., \& {et
		al}. 1999, \solphys, 187, 229
	
	\bibitem[{{Hood} \& {Priest}(1979)}]{hood1979}
	{Hood}, A.~W., \& {Priest}, E.~R. 1979, \solphys, 64, 303
	
	\bibitem[{{Hundhausen}(1987)}]{hundhausen1987}
	{Hundhausen}, A.~J. 1987, in Sixth International Solar Wind Conference, ed.
	V.~J. {Pizzo}, T.~{Holzer}, \& D.~G. {Sime}, 181
	
	\bibitem[{{Ji} {et~al.}(2003){Ji}, {Wang}, {Schmahl}, {Moon}, \&
		{Jiang}}]{haishengji2003}
	{Ji}, H., {Wang}, H., {Schmahl}, E.~J., {Moon}, Y.-J., \& {Jiang}, Y. 2003,
	\apjl, 595, L135
	
	\bibitem[{{Karpen} {et~al.}(2012){Karpen}, {Antiochos}, \&
		{DeVore}}]{karpen2012}
	{Karpen}, J.~T., {Antiochos}, S.~K., \& {DeVore}, C.~R. 2012, \apj, 760, 81
	
	\bibitem[{{Kippenhahn} \& {Schl{\"u}ter}(1957)}]{kippenhahn1957}
	{Kippenhahn}, R., \& {Schl{\"u}ter}, A. 1957, \zap, 43, 36
	
	\bibitem[{{Kliem} \& {T{\"o}r{\"o}k}(2006)}]{kliem2006}
	{Kliem}, B., \& {T{\"o}r{\"o}k}, T. 2006, Physical Review Letters, 96, 255002
	
	\bibitem[{{Kumar} {et~al.}(2012){Kumar}, {Cho}, {Bong}, {Park}, \&
		{Kim}}]{pankaj2012}
	{Kumar}, P., {Cho}, K.-S., {Bong}, S.-C., {Park}, S.-H., \& {Kim}, Y.~H. 2012,
	\apj, 746, 67
	
	\bibitem[{{Lemen} {et~al.}(2012){Lemen}, {Title}, {Akin}, {Boerner}, \& {et
			al}}]{lemen2012}
	{Lemen}, J.~R., {Title}, A.~M., {Akin}, D.~J., {Boerner}, P.~F., \& {et al}.
	2012, \solphys, 275, 17
	
	\bibitem[{{Lites}(2005)}]{lites2005}
	{Lites}, B.~W. 2005, \apj, 622, 1275
	
	\bibitem[{{Liu} {et~al.}(2007){Liu}, {Lee}, {Yurchyshyn}, {Deng}, {Cho},
		{Karlick{\'y}}, \& {Wang}}]{liuchang2007}
	{Liu}, C., {Lee}, J., {Yurchyshyn}, V., {Deng}, N., {Cho}, K.-s.,
	{Karlick{\'y}}, M., \& {Wang}, H. 2007, \apj, 669, 1372
	
	\bibitem[{{Liu} {et~al.}(2012){Liu}, {Kliem}, {T{\"o}r{\"o}k}, {Liu}, {Titov},
		{Lionello}, {Linker}, \& {Wang}}]{liur2012}
	{Liu}, R., {Kliem}, B., {T{\"o}r{\"o}k}, T., {Liu}, C., {Titov}, V.~S.,
	{Lionello}, R., {Linker}, J.~A., \& {Wang}, H. 2012, \apj, 756, 59
	
	\bibitem[{{Liu} {et~al.}(2010){Liu}, {Liu}, {Wang}, {Deng}, \&
		{Wang}}]{liur2010}
	{Liu}, R., {Liu}, C., {Wang}, S., {Deng}, N., \& {Wang}, H. 2010, \apjl, 725,
	L84
	
	\bibitem[{{Liu}(2008)}]{liuy2008}
	{Liu}, Y. 2008, \apjl, 679, L151
	
	\bibitem[{{Liu} {et~al.}(2009){Liu}, {Su}, {Xu}, {Lin}, {Shibata}, \&
		{Kurokawa}}]{liuy2009}
	{Liu}, Y., {Su}, J., {Xu}, Z., {Lin}, H., {Shibata}, K., \& {Kurokawa}, H.
	2009, \apjl, 696, L70
	
	\bibitem[{{Mackay} {et~al.}(2010){Mackay}, {Karpen}, {Ballester}, {Schmieder},
		\& {Aulanier}}]{mackay2010}
	{Mackay}, D.~H., {Karpen}, J.~T., {Ballester}, J.~L., {Schmieder}, B., \&
	{Aulanier}, G. 2010, \ssr, 151, 333
	
	\bibitem[{{Mari{\v c}i{\'c}} {et~al.}(2009){Mari{\v c}i{\'c}}, {Vr{\v s}nak},
		\& {Ro{\v s}a}}]{maricic2009}
	{Mari{\v c}i{\'c}}, D., {Vr{\v s}nak}, B., \& {Ro{\v s}a}, D. 2009, \solphys,
	260, 177
	
	\bibitem[{{Moore} \& {Sterling}(2006)}]{moore2006}
	{Moore}, R.~L., \& {Sterling}, A.~C. 2006, Washington DC American Geophysical
	Union Geophysical Monograph Series, 165, 43
	
	\bibitem[{{Moore} {et~al.}(2001){Moore}, {Sterling}, {Hudson}, \&
		{Lemen}}]{moore2001}
	{Moore}, R.~L., {Sterling}, A.~C., {Hudson}, H.~S., \& {Lemen}, J.~R. 2001,
	\apj, 552, 833
	
	\bibitem[{{Pesnell} {et~al.}(2012){Pesnell}, {Thompson}, \&
		{Chamberlin}}]{pesnell2012}
	{Pesnell}, W.~D., {Thompson}, B.~J., \& {Chamberlin}, P.~C. 2012, \solphys,
	275, 3
	
	\bibitem[{{Pneuman}(1983)}]{pneuman1983}
	{Pneuman}, G.~W. 1983, \solphys, 88, 219
	
	\bibitem[{{Priest} \& {Forbes}(2002)}]{priest2002}
	{Priest}, E.~R., \& {Forbes}, T.~G. 2002, \aapr, 10, 313
	
	\bibitem[{{Roussev} {et~al.}(2003){Roussev}, {Forbes}, {Gombosi}, {Sokolov},
		{DeZeeuw}, \& {Birn}}]{roussev2003}
	{Roussev}, I.~I., {Forbes}, T.~G., {Gombosi}, T.~I., {Sokolov}, I.~V.,
	{DeZeeuw}, D.~L., \& {Birn}, J. 2003, \apjl, 588, L45
	
	\bibitem[{{Rust} \& {Kumar}(1996)}]{rust1996}
	{Rust}, D.~M., \& {Kumar}, A. 1996, \apjl, 464, L199
	
	\bibitem[{{Schou} {et~al.}(2012){Schou}, {Scherrer}, {Bush}, {Wachter}, \& {et
			al}}]{schou2012}
	{Schou}, J., {Scherrer}, P.~H., {Bush}, R.~I., {Wachter}, R., \& {et al}. 2012,
	\solphys, 275, 229
	
	\bibitem[{{Schrijver} \& {De Rosa}(2003)}]{schrijver2003}
	{Schrijver}, C.~J., \& {De Rosa}, M.~L. 2003, \solphys, 212, 165
	
	\bibitem[{{Schuck}(2005)}]{schuck2005}
	{Schuck}, P.~W. 2005, \apjl, 632, L53
	
	\bibitem[{{Srivastava} {et~al.}(2010){Srivastava}, {Zaqarashvili}, {Kumar}, \&
		{Khodachenko}}]{srivastavaak2010}
	{Srivastava}, A.~K., {Zaqarashvili}, T.~V., {Kumar}, P., \& {Khodachenko},
	M.~L. 2010, \apj, 715, 292
	
	\bibitem[{{Su} \& {van Ballegooijen}(2013)}]{suyingna2013}
	{Su}, Y., \& {van Ballegooijen}, A. 2013, \apj, 764, 91
	
	\bibitem[{{Sun} {et~al.}(2014){Sun}, {Cheng}, \& {Ding}}]{sunjq2014}
	{Sun}, J.~Q., {Cheng}, X., \& {Ding}, M.~D. 2014, \apj, 786, 73
	
	\bibitem[{{Tandberg-Hanssen}(1998)}]{hanssen1998}
	{Tandberg-Hanssen}, E. 1998, in Astronomical Society of the Pacific Conference
	Series, Vol. 150, IAU Colloq. 167: New Perspectives on Solar Prominences, ed.
	D.~F. {Webb}, B.~{Schmieder}, \& D.~M. {Rust}, 11
	
	\bibitem[{{Tandberg-Hanssen} {et~al.}(1974){Tandberg-Hanssen}, {Hanssen}, \&
		{Riddle}}]{hanssen1974}
	{Tandberg-Hanssen}, E., {Hanssen}, R.~T., \& {Riddle}, A.~C. 1974, in \baas,
	Vol.~6, Bulletin of the American Astronomical Society, 295
	
	\bibitem[{{T{\"o}r{\"o}k} \& {Kliem}(2005)}]{torok2005}
	{T{\"o}r{\"o}k}, T., \& {Kliem}, B. 2005, \apjl, 630, L97
	
	\bibitem[{{T{\"o}r{\"o}k} {et~al.}(2004){T{\"o}r{\"o}k}, {Kliem}, \&
		{Titov}}]{torok2004}
	{T{\"o}r{\"o}k}, T., {Kliem}, B., \& {Titov}, V.~S. 2004, \aap, 413, L27
	
	\bibitem[{{van Ballegooijen} \& {Martens}(1989)}]{vanballegooijen1989}
	{van Ballegooijen}, A.~A., \& {Martens}, P.~C.~H. 1989, \apj, 343, 971
	
	\bibitem[{{Vemareddy} {et~al.}(2012{\natexlab{a}}){Vemareddy}, {Ambastha}, \&
		{Maurya}}]{vemareddy2012b}
	{Vemareddy}, P., {Ambastha}, A., \& {Maurya}, R.~A. 2012{\natexlab{a}}, \apj,
	761, 60
	
	\bibitem[{{Vemareddy} {et~al.}(2016){Vemareddy}, {Cheng}, \&
		{Ravindra}}]{vemareddy2016b}
	{Vemareddy}, P., {Cheng}, X., \& {Ravindra}, B. 2016, \apj, 829, 24
	
	\bibitem[{{Vemareddy} {et~al.}(2012{\natexlab{b}}){Vemareddy}, {Maurya}, \&
		{Ambastha}}]{vemareddy2012a}
	{Vemareddy}, P., {Maurya}, R.~A., \& {Ambastha}, A. 2012{\natexlab{b}},
	\solphys, 277, 337
	
	\bibitem[{{Vemareddy} \& {Zhang}(2014)}]{vemareddy2014b}
	{Vemareddy}, P., \& {Zhang}, J. 2014, \apj, 797, 80
	
	\bibitem[{{Vial} \& {Engvold}(2015)}]{VialEngvold2015}
	{Vial}, J.-C., \& {Engvold}, O., eds. 2015, Astrophysics and Space Science
	Library, Vol. 415, {Solar Prominences}
	
	\bibitem[{{Wang} {et~al.}(2017){Wang}, {Liu}, \& {Wang}}]{wangw2017}
	{Wang}, W., {Liu}, R., \& {Wang}, Y. 2017, \apj, 834, 38
	
	\bibitem[{{Wang} {et~al.}(2010){Wang}, {Cao}, {Chen}, {Zhang}, {Yu}, {Zheng},
		{Shen}, {Zhang}, \& {Wang}}]{wangyuming2010}
	{Wang}, Y., {Cao}, H., {Chen}, J., {Zhang}, T., {Yu}, S., {Zheng}, H., {Shen},
	C., {Zhang}, J., \& {Wang}, S. 2010, \apj, 717, 973
	
	\bibitem[{{Weber} {et~al.}(2004){Weber}, {Deluca}, {Golub}, \&
		{Sette}}]{weber2004}
	{Weber}, M.~A., {Deluca}, E.~E., {Golub}, L., \& {Sette}, A.~L. 2004, in IAU
	Symposium, Vol. 223, Multi-Wavelength Investigations of Solar Activity,
	321--328
	
	\bibitem[{{Wedemeyer-B{\"o}hm} {et~al.}(2012){Wedemeyer-B{\"o}hm}, {Scullion},
		{Steiner}, {Rouppe van der Voort}, {de La Cruz Rodriguez}, {Fedun}, \&
		{Erd{\'e}lyi}}]{wedemeyerbohm2012}
	{Wedemeyer-B{\"o}hm}, S., {Scullion}, E., {Steiner}, O., {Rouppe van der
		Voort}, L., {de La Cruz Rodriguez}, J., {Fedun}, V., \& {Erd{\'e}lyi}, R.
	2012, \nat, 486, 505
	
	\bibitem[{{Williams} {et~al.}(2005){Williams}, {T{\"o}r{\"o}k}, {D{\'e}moulin},
		{van Driel-Gesztelyi}, \& {Kliem}}]{williamsdr2005}
	{Williams}, D.~R., {T{\"o}r{\"o}k}, T., {D{\'e}moulin}, P., {van
		Driel-Gesztelyi}, L., \& {Kliem}, B. 2005, \apjl, 628, L163
	
	\bibitem[{{Yurchyshyn} {et~al.}(2006){Yurchyshyn}, {Karlick{\'y}}, {Hu}, \&
		{Wang}}]{yurchyshyn2006a}
	{Yurchyshyn}, V., {Karlick{\'y}}, M., {Hu}, Q., \& {Wang}, H. 2006, \solphys,
	235, 147
	
	\bibitem[{{Zhang} {et~al.}(2012){Zhang}, {Cheng}, \& {Ding}}]{zhangj2012}
	{Zhang}, J., {Cheng}, X., \& {Ding}, M.-D. 2012, Nature Communications, 3
	
\end{thebibliography}

\end{document}